\documentclass[amstex,12pt]{article}         
\usepackage{amssymb,amsmath}
\usepackage{epsf}
\newlength{\dinwidth}
\newlength{\dinmargin}
\newcommand{\ie}{{\it i.e.\ }}

\newcommand{\cf}{{\it cf.\ }}

\newcommand{\RR}{\mathbb R}
\newcommand{\CC}{\mathbb C}
\newcommand{\NN}{\mathbb N}

\newcommand{\state}[1]{{\langle \, #1 \, \rangle}}
\newcommand{\ob}[1]{{\langle \, #1 \, \rangle_\beta}}
\newcommand{\obt}[1]{{\langle \, #1 \, \rangle_\beta^T}}
\newcommand{\obd}[1]{{\langle \, #1 \, \rangle_{\beta,d}}}
\newcommand{\obf}[1]{{\langle \, #1 \, \rangle_{0,\beta}}}
\newcommand{\ovf}[1]{{\langle \, #1 \, \rangle_{0,\infty}}}

\newcommand{\bp}{{\mbox{\boldmath $p$}}}
\newcommand{\bk}{{\mbox{\boldmath $k$}}}
\newcommand{\bx}{{\mbox{\boldmath $x$}}}

\newcommand{\be}{{\mbox{\boldmath $e$}}}

\newcommand{\bO}{{\mbox{\boldmath $0$}}} 
\newcommand{\beps}{{\mbox{\boldmath $\varepsilon$}}}
\newcommand{\bfmu}{{\mbox{\footnotesize \boldmath $\mu$}}}
\newcommand{\bfp}{{\mbox{\footnotesize \boldmath $p$}}}
\newcommand{\bfx}{{\mbox{\footnotesize \boldmath $x$}}}

\newcommand{\bfe}{{\mbox{\footnotesize \boldmath $e$}}}

\newcommand{\fm}{{\mbox{\footnotesize $m$}}}
\newcommand{\fv}{{\mbox{\footnotesize $v$}}}
\newcommand{\fw}{{\mbox{\footnotesize $w$}}}
\newcommand{\fbeta}{{\mbox{\footnotesize $\beta$}}}
\newcommand{\fepsilon}{{\mbox{\footnotesize $\varepsilon$}}}
\newcommand{\fomega}{{\mbox{\footnotesize $\omega$}}}
\newcommand{\fsigma}{{\mbox{\footnotesize $\sigma$}}}
\newcommand{\fim}{{\mbox{\footnotesize $i$}}}
\newcommand{\fxo}{{\mbox{\footnotesize $x_0$}}}
\newcommand{\fpo}{{\mbox{\footnotesize $p_0$}}}

\newcommand{\wfm}{{W_{m,\beta}}}
\newcommand{\wfmh}{{h_{1,2}*W_{m,\beta}}}
\newcommand{\wfmho}{{h_{1,2}*W_{m_0,\beta}}}

\begin{document}
\title{\hspace*{-5mm} Asymptotic Dynamics of Thermal Quantum Fields}
\author{Jacques Bros$^a$ \hspace{0.5pt} and \hspace{0.5pt} Detlev Buchholz$^b$
\\[5mm]
${}^a \,$ Service de Physique Th\'eorique, CEA--Saclay\\
F--91191 Gif--sur--Yvette, France\\[2mm]
${}^b \,$ Institut f\"ur Theoretische Physik,
Universit\"at G\"ottingen\\
D--37073 G\"ottingen, Germany}
\date{}
\maketitle
%
%

\begin{abstract}
\noindent It is shown that
the timelike asymptotic properties of thermal correlation
functions in relativistic quantum field theory
can be described in terms of free fields
carrying some stochastic degree of freedom which couples to
the thermal background. These ``asymptotic thermal fields''
have specific algebraic properties (commutation relations) and their dynamics
can be expressed in terms of asymptotic field equations.
The formalism is applied to interacting theories where it
yields concrete non-perturbative results for the asymptotic
thermal propagators. The results are consistent with the
expected features of dissipative propagation of the
constituents of thermal states, outlined in previous work, and
they shed new light on the non-perturbative effects of thermal
backgrounds.
\end{abstract}

\section{Introduction}
\setcounter{equation}{0}
It is a basic difficulty in thermal quantum field theory
that the timelike asymptotic properties of thermal correlation
functions cannot be interpreted in terms of free fields due to
the omnipresent dissipative effects of the thermal background
\cite{NaReTh,La}. This well--known fact manifests itself in a
softened pole structure of the Green's functions in momentum
space and is at the root of the failure of the conventional
approaches to real-time thermal perturbation theory \cite{St,We,Abri}.

There exist several interesting proposals for the determination
of the asymptotic structure of thermal correlation functions
in Minkowski space and of their corresponding
momentum space properties, respectively. They range from
(a) the resummation of dominant low-energy self-energy diagrams,
such as the hard-thermal-loop approximation \cite{BrPi} (cf.\ 
also \cite{BlI,Lebe} and references quoted there),
through (b) the description of thermal propagators in terms of
generalized free fields with continuous mass spectrum
\cite{La} and (c) the idea of proceeding to effective theories, 
where the temperature dependence is incorporated into the
Lagrangian \cite{We}, up to (d) the passage to suitable macroscopic 
approximations, providing some insight into the asymptotic 
features of thermal propagators \cite{ArYa}. These various 
approaches have in common that they rely on {\it ad hoc}
assumptions based on physical considerations or experience with 
perturbative computations. Thus, what seems to be missing
is a more systematic analysis of the asymptotics of thermal quantum
fields, in analogy to collision theory in the vacuum theory.

We therefore reconsider in the present article the problem
of describing the asymptotic properties of thermal
quantum fields in a general framework set forth in 
\cite{BrBu2}. In this setting the thermal quantum fields
are described in terms of their $n$--point correlation
functions in the spirit of the Wightman approach to quantum
field theory in the vacuum sector. Fundamental features 
of the theory such as Einstein causality and the Gibbs-von Neumann
characterization of thermal equilibrium states can be expressed 
in terms of these functions in a simple and model independent
way. Moreover, these functions provide a convenient link between the various 
approaches to thermal quantum field theory which historically
have been developed independently, such as the real and
imaginary time formalisms and thermofield dynamics,
\cf \cite{LaWe} for a comprehensive review and an extensive
list of references.

Within this general framework we have established in \cite{BrBu2} a 
K\"all\'en--Lehmann type representation of the thermal 
two--point functions of interacting quantum fields which 
exhibits similar complex analytic structures and boundary
value relations on the reals as those found for free
fields in \cite{DoJa}. It turns out that this general
representation of two--point functions is particularly
well--suited to the study of the asymptotic properties
of thermal fields and their particle interpretation.
We shall exhibit this fact in the first part of the 
present analysis and establish a universal algebraic
setting for the discussion of these asymptotic structures.
This setting is compatible with the fundamental postulates of quantum
field theory and provides a basis for the analysis
of concrete models. In particular, 
it allows one to define a notion of asymptotic dynamics,
in accordance with the idea that the interaction of fields
must be taken into account in thermal states
also at asymptotic times. In the second part 
of our investigation we apply this
setting to some self-interacting theory and obtain concrete
results for the corresponding asymptotic propagators
which go beyond perturbation theory.

Our article is organized as follows.
In Sec.\ 2 we recall the general properties of
thermal quantum field theory and determine the
asymptotic structure in the (real) time 
coordinates of the thermal correlation functions
under the assumption that the thermal states are
composed of massive constituents and that
collective memory effects are asymptotically sub-dominant.
Section 3 contains the definition and analysis
of the asymptotic thermal fields. It is shown that
these fields reproduce the asymptotic structure
of the correlation functions and that their normal products
can consistently be defined. These products are the
basic ingredient in
the asymptotic dynamics. The formalism is applied
to a self--interacting scalar field in Sec.~4 and the
paper concludes with a brief discussion of the results.
Some technical points are deferred to two appendices.

\section{Asymptotic thermal correlation functions}
\setcounter{equation}{0}
In this section we determine the
timelike asymptotic structure of thermal correlation functions
in the setting of (real-time) thermal quantum field theory,
starting from some quite general assumptions.
In order to simplify this discussion, we restrict our
attention to theories of a real scalar
field $\phi$. But it will
become clear that our approach can be extended to more complex
situations with only little more effort. We begin by recalling
some relevant notions in order to fix our notation.

The basic object in our analysis is the field $\phi (x)$ which,
technically speaking, is an operator-valued distribution, \ie the
averaged expressions
\begin{equation} \label{e.smear}
\phi (f) \doteq \int \! dx \, f(x) \phi (x),
\end{equation}
where $f$ is any test function with compact
support in Minkowski space $\RR^4$,
are defined on some common dense and stable domain in the
underlying Hilbert space of states.
The finite sums and products of the averaged field operators
generate a *-algebra $\cal{A}$
consisting of all polynomials $A$ of the form
\begin{equation} \label{polynom}
A=\sum \phi (f_{1}) \phi (f_{2})\cdots \phi (f_{n}).
\end{equation}
We assume that also all multiples of the unit operator $1$ are
elements of ${\cal A}$. As the field $\phi$ is real, the
*-operation (Hermitian conjugation) on $\cal{A}$ is defined by
\begin{equation} \label{involution}
A^{\ast } \doteq \sum \phi(\overline{f_{n}}) \cdots
\phi (\overline{f_{2}}) \phi (\overline{f_{1}}),
\end{equation}
where $\overline{f}$ denotes the complex conjugate of $f$.
The translations $x \in \RR^4$ act on $\cal{A}$
by automorphisms $\alpha_x$,
\begin{equation} \label{e.transl}
\alpha_x (A) \doteq
\sum \phi (f_{1,x}) \phi (f_{2,x})\cdots \phi (f_{n,x}),
\end{equation}
where $f_x$ is defined by
$f_x (y) \doteq f(y - x)$, $y \in \RR^4$.
As we restrict attention here to scalar fields, we also have
the causal commutation relations (locality)
\begin{equation}
[\phi(f_1), \phi(f_2)] = 0
\end{equation}
if the supports of $f_1$ and $f_2$ are spacelike separated.

We mention as an aside that
in the real-time formalism of thermal field theory
it is frequently convenient to introduce a second field
(the so--called tilde field) which is anti--isomorphic to the
basic field and commutes with it everywhere. One then deals
with an irreducible set of field operators, similarly to
the vacuum theory. However, we do not need to make use of this
extension of the formalism here.

The states (expectation functionals) on the algebra $\cal{A}$,
generically denoted~by $\state{\cdot}$, have the defining properties
$\state{c_{1}A_{1}+c_{2}A_{2}} = c_{1} \state{A_{1}} +c_{2}
\state{A_{2}} $, $c_1,c_2 \in \CC$  (linearity),
$\state{A^*A} \geq 0$ (positivity), and
$\state{1} =1$ (normalization). Fixing a Lorentz frame
with corresponding space-time coordinates $x = (x_0, \bx)$, the thermal
equilibrium states in that frame can be distinguished by the KMS
condition. \\[2mm]
{\bf KMS condition:} A state $\ob{\cdot}$ satisfies
the KMS condition at inverse temperature $\beta > 0$
if for each pair of operators
$A,A^\prime \in {\cal A}$ there is some function $F$ which is analytic
in the strip $S_\beta \doteq \{ z \in \CC : 0 <
\mbox{Im} z < \beta \} $
and continuous at the boundaries such that
\begin{equation}
F(x_0) = \ob{A^\prime \alpha_{x_0}(A)},
\ \ F(x_0+i\beta) = \ob{\alpha_{x_0}(A) A^\prime},
\quad x_0 \in \RR,
\end{equation}
where $\alpha_{x_0}$ denotes the action of the time translations.
In this situation,
$\ob{\cdot}$ is called a KMS state.\\[2mm]
\indent We will also make use of a slightly stronger version
of the KMS condition, proposed in \cite{BrBu1}. It can be established
under natural conditions on the underlying theory and is a remnant of
the relativistic spectrum condition in the vacuum sector. It was
therefore called relativistic KMS condition in \cite{BrBu1}. \\[2mm]
{\bf Relativistic KMS condition:} A state $\ob{\cdot}$ is said
to satisfy the relativistic KMS condition at inverse temperature
$\beta > 0$ if for each pair of operators
$A,A^\prime \in {\cal A}$ there is some function $F$ which is analytic in the
tube $\RR^4 + i \,\big( V_+\cap((\beta,\bO) -  V_+) \big)$ such that in the
sense of continuous boundary values \cite{BrBu1}
\begin{equation}
F(x) = \ob{A^\prime \alpha_x(A)}, \ \
F(x + i (\beta,\bO)) = \ob{\alpha_x(A) A^\prime}, \quad x \in \RR^4.
\end{equation}

\indent We restrict attention here to
KMS states $\ob{\cdot}$ which are homogeneous and isotropic
in their rest systems and describe pure phases. It follows from
the latter assumption that the corresponding correlation
functions have timelike clustering properties
(\ie they are weakly mixing).
Hence we may assume without loss of generality that
$\ob{\phi(f)} = 0$, subtracting from the field
$\phi$ some constant, if necessary.

The preceding general assumptions imply that there holds a
K\"allen-Lehmann type representation for the two-point functions
of the field $\phi(x)$ in all KMS states \cite{BrBu2}.
We note that we will deal in the subsequent analysis with the
unregularized fields in order to simplify the notation;
the following statements are thus to be understood in the sense
of temperate distributions. The two-point functions
can be represented in the form \cite{BrBu2}
\begin{equation} \label{rep}
\ob{\phi(x_1) \phi(x_2)}
= \int_0^\infty \! dm \, D_\beta (m, \bx_1 - \bx_2) \,
\wfm (x_1 - x_2),
\end{equation}
where $\wfm (x)$ is the two-point function of a free
scalar field of mass $m$ for a KMS state of inverse temperature $\beta$,
\begin{equation} \label{free}
\wfm (x) = (2\pi)^{-3} \int \! dp \, \varepsilon (p_0) \delta (p^2 - m^2)
\, (1 - e^{-\beta p_0})^{-1} e^{-ipx}.
\end{equation}
The specific features of the underlying theory are contained in
the distribution $D_\beta (m,\bx)$,
called damping factor in \cite{BrBu2,BrBu3}, which
is regular (as a matter of fact:~analytic) in $\bx$ if the underlying
KMS state satisfies the relativistic KMS condition \cite{BrBu2}.
Moreover, it is rotational invariant because of the isotropy of the state.

The familiar K\"allen-Lehmann representation of the
vacuum state can be recovered from (\ref{rep}) in
the limit $\beta \rightarrow \infty$.
There $D_\infty (m,\bx)$
does not depend on $\bx$ and is (the density of) a temperate measure with
respect to $m$.
We expect that the latter feature continues to hold at finite
temperatures in generic cases,
\ie the worst singularities which can appear in $D_\beta (m,\bx)$
with respect to $m$ ought to be $\delta$-functions. In order to simplify
the subsequent analysis, we make here the more specific assumption that
$D_\beta (m,\bx)$ can be decomposed into a discrete and
an absolutely continuous part,
\begin{equation}
D_\beta (m,\bx) = \delta (m - m_0) \, D_{\beta,d} (\bx) +
D_{\beta, c} (m,\bx),
\end{equation}
where $m_0 > 0$ is some fixed mass and
$D_{\beta, c} (m,\bx) (1 + m)^{-N}$ is, for sufficiently large $N$,
absolutely integrable with respect to $m$, the integral
being uniformly bounded for $\bx$ varying in compact sets.
Situations where the damping factors
contain several $\delta$-contributions or where the remainder
$D_{\beta, c} (m,\bx)$ exhibits a less regular behavior can be
treated with some more effort.

It was argued in \cite{BrBu3} that the $\delta$-contributions in the
damping factors are due to stable constituent particles of mass $m_0$
out of which the thermal states are formed, whereas the
collective quasiparticle-like excitations only contribute to the
continuous part of the damping factors. This assertion
will be substantiated by our present results.

Since the time dependence of the two-point functions is entirely
contained in the explicitly known functions $\wfm$ in the
representation (\ref{rep}), it is possible to analyze
their timelike asymptotic structure in the present general setting. We will
see that, disregarding low energy excitations, the asymptotically
leading terms are due to the $\delta$-contributions in $D_\beta$.
The low energy excitations can be suppressed
by regularizing the field $\phi(x)$ with respect
to the time variable with some test function $g$ whose Fourier transform
vanishes at the origin,
\begin{equation} \label{convolution}
\phi_g (x) = \int \! dy_0 \, g(y_0) \, \phi (x_0 + y_0,\bx).
\end{equation}
We mention as an aside that these partially regularized fields
already define (unbounded) operators if there holds
the relativistic KMS condition.
Plugging these regularized fields into (\ref{rep}), we obtain
\begin{equation} \label{rep2}
\ob{\phi_{g_1} (x_1) \phi_{g_2} (x_2)}
= \int_0^\infty \! dm \, D_\beta (m, \bx_1 - \bx_2) \
\wfmh (x_1 - x_2),
\end{equation}
where $h_{1,2} = g_1*g_2$ and the star denotes convolution with
respect to the time variable.
Hence only the function $\wfm$ is affected by the
regularization. The form of
$\wfmh (x)$ at asymptotic times $x_0$ and its dependence on $m$
are determined in Appendix A. One obtains
\begin{equation} \label{freetwopoint}
\wfmh (x)
= x_0^{-3/2} \big( e^{-imx_0} k_+(m) +
e^{imx_0} k_-(m) \big) + r(m,x),
\end{equation}
where $k_\pm (m)$ are continuous functions which decrease
faster than any inverse power of $m$ in the limit of large $m$.
The remainder $r(m,x)$ is continuous in $m$ and,
for $0 < \delta < 1/2$, bounded by
\begin{equation}
|r(m,x)| \leq c_N \, x_0^{-3/2 - \delta} \,  (1 + m)^{-N},
\quad N \in \NN,
\end{equation}
uniformly in $\bx$ on compact sets ${\cal C}$. Thus, in view of the
Riemann-Lebesgue theorem and the anticipated integrability properties of
$D_{\beta, c} (m,\bx)$, there holds
\begin{equation} \label{asympk} 
\lim_{x_0 \rightarrow \pm \infty} \int_0^\infty \! dm \,
D_{\beta, c} (m,\bx) \, e^{\mp imx_0} \, k_\pm (m) = 0,
\end{equation}
and the bound on $r(m,x)$ implies
\begin{equation}  \label{asympr} 
\lim_{x_0 \rightarrow \pm \infty} |x_0|^{3/2}
\int_0^\infty \! dm \, D_{\beta, c} (m,\bx) \, r(m,x) = 0.
\end{equation}
Both limits exist uniformly in $\bx \in {\cal C}$.
Setting
\begin{equation} \label{twopoint}
\obd{\phi_{g_1}(x_1) \phi_{g_2}(x_2)}
= D_{\beta,d}(\bx_1 - \bx_2) \ \wfmho (x_1 - x_2)
\end{equation}
and $\Delta_0 = |{x_{1}}_0 - {x_{2}}_0|$, we thus obtain for 
asymptotic $\Delta_0$
\begin{equation} \label{asymp}
\ob{\phi_{g_1}(x_1) \phi_{g_2}(x_2)}
= \obd{\phi_{g_1}(x_1) \phi_{g_2}(x_2)},
\end{equation}
disregarding terms which which decrease more rapidly than 
$\Delta_0^{-3/2}$ uniformly for $\bx_1,\bx_2$ varying in compact sets.

So if discrete parts are present in $D_\beta$,  
we see from relations (\ref{freetwopoint}) to (\ref{twopoint})
that the two-point functions exhibit  
for large timelike separations $\Delta_0$ of the fields 
a $\Delta_0^{-3/2}-$type behavior.
The continuous part of $D_\beta$ 
definitely gives rise to a more rapid decay. A leading
$\Delta_0^{-3/2}$ behavior of the two-point functions has indeed 
been established in some interacting models
\cite{ArYa,We2}; we are interested here in these asymptotically
dominant contributions.

Much less is known about the asymptotic properties of the
higher $n$-point functions and we therefore have to rely
on some {\it ad hoc} assumption which is expected to
cover a large class of physically interesting cases.
Namely, we assume that the timelike asymptotic behavior
of the $n$-point functions is governed by their
disconnected parts.
More precisely, if $\Delta_0 = \inf \, \{ |x_{j \, 0} - x_{k \, 0}| :
j,k=1,\dots n, \, j \neq k \},$
we anticipate that
there holds for the truncated (connected) $n$-point functions,
in which the fields are regularized as before,
\begin{equation} \label{clustering}
\lim_{\Delta_0 \rightarrow \infty} \Delta_0^{3(n-1)/2 - \delta} \,
\obt{\phi_{g_{1}}(x_{1}) \cdots \phi_{g_{n}}(x_{n})} = 0, \quad
\delta > 0,
\end{equation}
where $T$ denotes truncation. Since the low-energy contributions
are suppressed in these functions, their dominant asymptotic
behavior should again be due to the exchange of constituent
particles between the space-time positions of all fields,
leading to  (\ref{clustering}).
In heuristic terms, this assumption means that there are no
collective memory effects present in the underlying KMS states.
This situation may be expected
to prevail at sufficiently high temperatures.
With this input we obtain, by decomposing the $n$-point functions
into sums of products of truncated functions (cluster decomposition),
taking into account the preceding results on the two-point function
and the fact that the one-point function vanishes,
\begin{equation}
\begin{split}
& \lim_{\Delta_0 \rightarrow \infty} \Delta_0^{3n/4} \,
\big| \ob{\phi_{g_{1}}(x_{1}) \cdots \phi_{g_{n}}(x_{n})} \\
& - \sum_{\mbox{\tiny partitions}}
\ob{\phi_{g_{i_1}} (x_{i_1}) \phi_{g_{i_2}} (x_{i_2})}
\cdots
\ob{\phi_{g_{i_{(n-1)}}} (x_{i_{(n-1)}}) \phi_{g_{i_n}} (x_{i_n})}
\big| = 0,
\end{split}
\end{equation}
where the sum extends over all ordered partitions
of $\{1,\dots n \}$ if $n$ is even and is to be replaced by $0$
if $n$ is odd. As the asymptotic behaviour of the two-point
functions in turn is given by (\ref{asymp}), we arrive at
an almost explicit description of the asymptotically leading
contributions to the $n$-point functions:
disregarding terms
which decay more rapidly than $\Delta_0^{-3n/4}$, it is given by
\begin{equation}
\begin{split}
& \ob{\phi_{g_{1}}(x_{1}) \cdots \phi_{g_{n}}(x_{n})} \\
& = \sum_{\mbox{\tiny partitions}}
\obd{\phi_{g_{i_1}} (x_{i_1}) \phi_{g_{i_2}} (x_{i_2})}
\cdots
\obd{\phi_{g_{i_{(n-1)}}} (x_{i_{(n-1)}}) \phi_{g_{i_n}} (x_{i_n})}
\end{split}
\end{equation}
if $n$ is even and $0$ if $n$ is odd. So the only unknown
is the discrete part $D_{\beta,d}$ of the damping factor
entering into $\obd{\, \cdot \,}$.
In the subsequent section we will show that these leading
contributions can be described in terms of a new type of asymptotic
field. This insight will allow us to determine
$D_{\beta,d}$ in concrete models.

\section{Asymptotic thermal fields and their dynamics}
\setcounter{equation}{0}
We have determined in the preceding section the asymptotically
leading contributions of the $n$-point functions under quite
general assumptions. The upshot of this analysis is the insight
that these contributions have the structure of quasi-free states,
\ie they can be expressed in terms of sums of products of
two-point functions. One could interpret this
structure in terms of generalized free fields having $c$-number
commutation relations \cite{La}. We do not rely here on this
idea, however, since these commutation relations would
depend on the specific features of the underlying theory and on
the temperature of the states which both
enter into the damping factors. Instead, we propose a 
model-independent setting which allows one to describe these
structures in terms of a universal quantum
field which admits KMS states of arbitrary temperatures and
with arbitrary damping factors. Moreover, one can
define normal products of this field with respect to the
vacuum state which are of relevance in the formulation
of the asymptotic dynamics.

The basic idea in our approach is to proceed from the familiar
free scalar massive field to a suitable 
central extension.
The resulting hermitian field is denoted by $\phi_0$. It is,
in the states of interest here, an operator-valued
distribution and satisfies in the distinguished
Lorentz frame the commutation relations
\begin{equation} \label{commutator}
[\phi_0 (x_1), \phi_0 (x_2)] = \Delta_{m_0} (x_1 - x_2) \, Z(\bx_1 - \bx_2),
\end{equation}
where $\Delta_{m_0}$ is the familiar commutator function of
a free scalar field of mass $m_0$,
\begin{equation} \label{paulijordan}
\Delta_{m_0} (x) = (2\pi)^{-3} \int \! dp \, \varepsilon (p_0)
\delta (p^2 - m_0^2) \, e^{-ipx},
\end{equation}
and $Z$
is an operator-valued distribution commuting with $\phi_0$
(and hence with itself). In order to make these
relations algebraically consistent, we have to assume
\begin{equation}
Z(-\bx) = Z(\bx)= Z(\bx)^*.
\end{equation}
We note that the right hand side of relation (\ref{commutator}) is
well defined in the sense of distributions since $\Delta_{m_0}$
becomes a test function in the spatial variables after
regularization with respect to the time variable. The action
of the space-time translations on $\phi_0$ and
$Z$ can consistently be defined by
\begin{equation}
\alpha_y(\phi_0(x)) = \phi_0(x+y), \quad \alpha_y(Z(\bx)) = Z(\bx),
\quad y \in \RR^4.
\end{equation}
Hence $\phi_0$ is a local field which transforms covariantly
under space-time translations and therefore fits into the
general setting outlined in Sec.\ 2.
In particular, one can proceed from
the field $\phi_0$ to a polynomial *-algebra ${\cal A}_0$ which
is generated by all finite sums of products of the
averaged field operators.
We emphasize that we do not impose from the outset
any field equations on $\phi_0$ since this would
reduce the flexibility of the formalism which is
designed to cover a large class of theories with different
dynamics.

Let us determine next the KMS states $\obf{\cdot}$ on the
algebra ${\cal A}_0$. Again,
we restrict attention to states which are isotropic
in the distinguished Lorentz frame and
have timelike clustering properties.
The computation of the $n$-point functions of the
KMS states can then be performed by standard arguments.

As before,
we may assume without loss of generality that $\obf{\phi_0(x)} = 0$.
Moreover, since the time dependence of the commutator (\ref{commutator})
is entirely contained in the distribution $\Delta_{m_0}$, we obtain
for pure time translations $y = (y_0,0)$ because of the
timelike clustering properties of the states
\begin{equation}
\begin{split}
& \obf{\phi_0 (x_1) \cdots \phi_0 (x_{n-2}) \,
[\phi_0 (x_{n-1}), \phi_0 (x_n)] } \\[4pt]
& = \lim_{y_0 \rightarrow \infty}
\obf{\phi_0 (x_1) \cdots \phi_0 (x_{n-2}) \,
[\phi_0 (x_{n-1} + y), \phi_0 (x_n + y)] } \\
& = \obf{\phi_0(x_1) \cdots \phi_0 (x_{n-2})} \,
\obf{ [\phi_0 (x_{n-1}), \phi_0 (x_n) ] },
\end{split}
\end{equation}
which implies
\begin{equation}
\begin{split}
& \obf{\phi_0(x_1) \cdots \phi_0 (x_{n-2}) \, Z(\bx_{n-1} - \bx_n)} \\
& = \obf{\phi_0(x_1) \cdots \phi_0 (x_{n-2})} \,
\obf{Z(\bx_{n-1} - \bx_n)}.
\end{split}
\end{equation}
Thus the operators $Z(\bx)$ can be replaced by $\obf{Z(\bx)} \cdot 1$
everywhere in the $n$-point functions.

Denoting by $\widehat{F}(\omega,\bx) = (2 \pi)^{-1/2}
\int \! dx_0 \, e^{-i x_0 \omega} \, F(x_0,\bx)$ the partial Fourier
transform of (operator-valued) distributions $F$ with respect
to the time variable, we get, by applying the KMS condition to the
$n$-point functions,
\begin{equation}
\begin{split}
& (1 - e^{- \beta \omega}) \, \obf{\phi_0(x_1) \cdots \phi_0(x_{n-1}) \,
\widehat{\phi_0} (\omega, \bx_n)} \\
& = \obf{\phi_0(x_1) \cdots \phi_0(x_{n-1}) \,
\widehat{\phi_0} (\omega, \bx_n) } -
\obf{\widehat{\phi_0} (\omega, \bx_n) \phi_0(x_1)
\cdots \phi_0(x_{(n-1)})}.
\end{split}
\end{equation}
Commuting $\widehat{\phi_0} (\omega, \bx_n)$ in the latter
term to the right and making use of
(\ref{commutator}) as well as of
the preceding results, we can proceed further to
\begin{eqnarray}
\hspace*{-12pt}
& = & \sum_{j=1}^{n-1} \obf{\phi_0(x_1) \cdots Z(\bx_j - \bx_n)
\cdots \phi_0(x_{n-1})} \,
e^{-i x_{j_0} \omega} \,
\widehat{\Delta_{m_0}} (\omega, \bx_j - \bx_n ) \\
& = &  \sum_{j=1}^{n-1}
\obf{\phi_0(x_1) \cdots \overset{j}{\vee}
\cdots \phi_0(x_{n-1}) \, \overset{n}{\vee}} \,
e^{-i x_{j_0} \omega} \,
\widehat{\Delta_{m_0}} (\omega, \bx_j - \bx_n ) \, \obf{Z(\bx_j -
  \bx_n)}, \nonumber
\end{eqnarray}
where the symbol $\overset{\cdot}{\vee}$ denotes omission of
the respective field. This relation can be divided without
ambiguities by $(1 - e^{- \beta \omega})$;
for $\obf{\phi_0(x_1) \cdots \phi_0(x_{(n-1)}) \,
\widehat{\phi_0} (\omega, \bx_n)}$ is (the density of) a
complex measure with respect to
$\omega$ which does not contain a discrete part at $\omega = 0$
because of the time-like clustering property
of $\obf{\cdot}$ and $\obf{\phi_0(x)} = 0$.
Performing the inverse Fourier transform of the
resulting expression with respect to $\omega$, we thus arrive at
\begin{eqnarray} \label{formula}
& & \obf{\phi_0(x_1) \cdots \phi_0(x_{n-1}) \, \phi_0 (x_n)} \\
& = & \sum_{j=1}^{n-1}
\obf{\phi_0(x_1) \cdots \overset{j}{\vee} \cdots
\phi_0(x_{n-1}) \, \overset{n}{\vee} } \, \obf{Z(\bx_j - \bx_n)} \,
W_{m_0,\beta} (x_j - x_n), \nonumber
\end{eqnarray}
where $W_{m_0, \beta}$ is the thermal two-point function
of a free field of mass $m_0$, defined in (\ref{free}).
As this relation holds for any $n$, we find that each
$\obf{\cdot}$ is a quasifree state with vanishing one-point
function and two-point function given by
\begin{equation} \label{freekms}
\obf{\phi_0(x_1) \phi_0(x_2)} = \obf{Z(\bx_1- \bx_2)} \,
W_{m_0,\beta} (x_1 - x_2),
\end{equation}
all higher truncated $n$-point functions being zero.
The factor $\obf{Z(\bx_1- \bx_2)}$ is not fixed by the KMS condition,
so the set of KMS states on ${\cal A}_0$ is degenerate
for any given temperature. KMS states corresponding to different
functions $\obf{Z(\bx_1- \bx_2)}$ lead to disjoint
representations of ${\cal A}_0$, describing macroscopically different
physical situations.

By a similar argument one can also determine the vacuum state
on ${\cal A}_0$ which is distinguished by the fact that
it is invariant under space-time translations and
complies with the relativistic spectrum condition.
As we do not deal here
with any other states of zero temperature, we denote the vacuum
state by $\ovf{\cdot}$ without danger of confusion. This state is
also quasifree and has a
vanishing one-point function and a two-point function given by
\begin{equation} \label{vacuum}
\ovf{\phi_0(x_1) \phi_0(x_2)} = \ovf{Z(0)} \,
W_{m_0,\infty} (x_1 - x_2).
\end{equation}
Note that a non-trivial $\bx$-dependence of $\ovf{Z(\bx)}$
would be incompatible with the relativistic
spectrum condition and the positivity of the state.
So this function must be constant and one may normalize the field
such that $\ovf{Z(0)} = 1$.
With this normalization, the vacuum representation
of $\phi_0$ is unique and coincides with the familiar
Fock-representation of a free scalar field of mass $m_0$.

Let us compare these results with the asymptotic structure
of the thermal correlation functions, established in the preceding section.
Identifying $\obf{Z(\bx_1- \bx_2)}$
with $D_{\beta,d}(\bx_1 - \bx_2)$ in relation (\ref{twopoint}), we
see that the expectation values of the
field $\phi_0$ in the KMS states
reproduce the asymptotically leading contributions of
the thermal correlation functions. Thus the possible
asymptotic structure of the thermal correlation
functions is encoded in the algebraic properties
of the field $\phi_0$, and this justifies its
interpretation as asymptotic thermal field.

We turn next to the definition of normal products
of the field $\phi_0$ which enter into the
asymptotic dynamics. For the sake of simplicity, we
will restrict attention here to those cases, where
the functions $\obf{Z(\bx)}$ are regular in $\bx$.
The case of singular functions requires more work and
will be discussed in
the subsequent section. As in the familiar case
of ordinary free fields, the normal products can be
defined by point splitting. Introducing the time
translations $\varepsilon = (\varepsilon,0)$, $\varepsilon \neq 0$,
the normal-ordered square of the field is given by
\begin{equation}
\phi_0^2 (x) \doteq \lim_{\varepsilon \rightarrow 0} \,
\big( \phi_0(x+\varepsilon) \phi_0(x-\varepsilon)
- \ovf{\phi_0(x+\varepsilon) \phi_0(x-\varepsilon)} \, 1 \big),
\end{equation}
and its normal-ordered cube is defined by
\begin{equation}
\begin{split}
& \lefteqn{\phi_0^3 (x) \doteq
\lim_{\varepsilon \rightarrow 0} \,
\big( \phi_0(x+\varepsilon) \phi_0(x) \phi_0(x-\varepsilon)
- \ovf{\phi_0(x+\varepsilon) \, \phi_0(x)} \,
\phi_0 (x-\varepsilon)}  \\
& -  \ovf{\phi_0(x+\varepsilon) \phi_0(x-\varepsilon)} \,
\phi_0 (x)
- \ovf{\phi_0(x) \phi_0(x-\varepsilon)} \,
\phi_0 (x+\varepsilon) \big).
\end{split}
\end{equation}
In a similar manner one can define higher normal products of the field.
They are well defined in all KMS states on
${\cal A}_0$, provided $\obf{Z(0)} = 1$. We will prove this
for the above examples.
Relations (\ref{freekms}) and (\ref{vacuum}) imply, because of
the time independence and normalization of $Z$
that, for $\varepsilon_1 \neq \varepsilon_2$,
\begin{equation}
\begin{split}
& \obf{\phi_0(x+ \varepsilon_1)  \phi_0(x+\varepsilon_2)} -
\ovf{\phi_0(x+\varepsilon_1) \phi_0(x+\varepsilon_2)} \\[6pt]
& = W_{m_0,\beta}(\varepsilon_1 - \varepsilon_2) -
W_{m_0,\infty}(\varepsilon_1 - \varepsilon_2)
\\
& = 2 \, (2 \pi)^{-3} \int \! dp \ \theta (p_0) \delta (p^2 - m_0^2) \,
(e^{\beta p_0} - 1)^{-1} \, \cos((\varepsilon_1 - \varepsilon_2) p_0).
\end{split}
\end{equation}
So this expression is well-defined and has,
for $\varepsilon_1,\varepsilon_2 \rightarrow 0$,
the limit
\begin{equation} \label{thermalfactor}
k (\beta) = 2 \,
(2 \pi)^{-3} \int \! dp \ \theta (p_0) \delta (p^2 - m_0^2) \,
(e^{\beta p_0} - 1)^{-1}.
\end{equation}
Making use of the fact that the thermal correlation functions of the
quasifree states $\obf{\cdot}$
can be decomposed into sums of products of two-point functions,
one gets for the non-trivial
$n$-point functions
involving $\phi_0^2$
\begin{equation} \label{square}
\begin{split}
& \obf{\phi_0(x_1) \cdots \phi_0(x_{j-1}) \, \phi_0^2(x_j) \,
\phi_0(x_{j+1}) \cdots \phi_0(x_n)} \\[6pt]
& = k(\beta) \, \obf{\phi_0(x_1) \cdots \phi_0(x_{j-1})
\phi_0(x_{j+1}) \cdots \phi_0(x_n)} \\
& + {\sum_{\mbox{\tiny partitions}}}^{\! \prime}
\obf{\phi_0(x_{j_1}) \phi_0(x_{j_2})} \cdots
\obf{\phi_0(x_{j_{n}}) \phi_0(x_{j_{n+1}})},
\end{split}
\end{equation}
where the primed sum extends over all partitions of
$\{1, \dots n \}$ into ordered pairs of mutually
different numbers, taking into account twice the index $j$ of
the normal-ordered square.
Similarly, one obtains for the $n$-point functions
involving $\phi_0^3$
\begin{equation} \label{cube}
\begin{split}
& \obf{\phi_0(x_1) \cdots \phi_0(x_{j-1}) \, \phi_0^3(x_j) \,
\phi_0(x_{j+1}) \cdots \phi_0(x_n)} \\[6pt]
& = 3 \, k(\beta) \, \obf{\phi_0(x_1) \cdots \phi_0(x_{j-1}) \phi_0(x_j)
\phi_0(x_{j+1}) \cdots \phi_0(x_n)} \\
& + {\sum_{\mbox{\tiny partitions}}}^{\! \prime}
\obf{\phi_0(x_{j_1}) \phi_0(x_{j_2})} \cdots
\obf{\phi_0(x_{j_{n+1}}) \phi_0(x_{j_{n+2}})},
\end{split}
\end{equation}
where the index $j$ of the normal-ordered cube now appears
three times in the partitions entering in the
primed sum. We refrain from giving here the
corresponding formulas for arbitrary normal-ordered powers.

It is important to notice that the timelike asymptotic
behaviour of the normal-ordered products
in KMS states differs from that in the vacuum due
to the appearance of the terms involving the factor $k(\beta)$.
In the case of the square, the asymptotically leading
contribution is $k(\beta) \, 1$.
This contribution can be suppressed
by convolution with a test function as in
(\ref{convolution}). The asymptotically dominant
contribution in the case of the cube is given
by $3 \, k(\beta) \, \phi_0(x_j)$, so this term
exhibits the same behaviour as the field
and can therefore not be neglected even if its
low energy components are suppressed.

After these preparations we can now turn to the
problem of formulating a dynamical law for the
asymptotic thermal field $\phi_0$. To this end, let
us assume that the underlying interacting field
satisfies some field equation. As $\phi_0$
describes the interacting field only asymptotically,
one may not expect that $\phi_0$ satisfies this
equation, too. But it should satisfy it in an asymptotic sense
which is consistent with our asymptotic approximation.
More concretely, let us consider the expression
\begin{equation} \label{fieldequation}
N(x) = \square \, \phi_0 (x) + m_0^2 \, \phi_0 (x)
+ \sum_{k=2}^K c_k \, \phi_0^k (x),
\end{equation}
where $c_k$ are fixed, state-independent (and hence
temperature-independent) constants. The
relation $N(x) = 0$ would say that $\phi_0$
satisfies a field equation for given mass
$m_0$ and coupling constants $c_k$. Because of
the commutation relations of $\phi_0$, this is clearly impossible
if one of the $c_k$ is different from zero. But, taking
into account the preceding remarks, it is sufficient to
demand that the operator $N(x)$ is asymptotically
negligible. More precisely, all
matrix elements of $N(x)$ should, after suppression
of the zero-energy contributions, decay more rapidly
at asymptotic times than those of the
field $\phi_0 (x)$ and its normal-ordered products,
\ie $ x_0^{3/2} \, N_g(x) \rightarrow  0$ as $x_0 \rightarrow \infty$.

It is a fundamental fact that this condition
cannot be satisfied by all KMS states on ${\cal A}_0$.
This feature is at the origin of the failure of the
conventional approaches to thermal perturbation theory,
where one starts from KMS states in free field theory
which are incompatible
with the underlying asymptotic dynamics. As a matter of fact,
the asymptotically admissible states are fixed by
$N(x)$, as we shall see. \\[2mm]
{\bf Definition:} Let $N(x)$ be given. A KMS state is said
to be  compatible with the asymptotic field equation
$ x_0^{3/2} \, N_g(x) \rightarrow 0$ if
\begin{equation} \label{constraint}
\lim_{x_0 \rightarrow \infty}
x_0^{3/2} \, \obf{\phi_0(x_1) \cdots \phi_0(x_{j}) \, N_g(x) \,
\phi_0(x_{j+1}) \cdots \phi_0(x_n)} = 0,
\end{equation}
for all $j,n \in \NN_0$, where $N_g(x)$ is regularized as in
(\ref{convolution}).
\\[2mm]
The vacuum complies with
this compatibility condition for any choice of $N(x)$ since
$(\square \, + m_0^2) \phi_0(x) = 0$ in the vacuum state and
$k(\infty) = 0$, \cf (\ref{thermalfactor}). KMS states
satisfy the condition if and only if it holds for
$n=1$; for the even powers of $\phi_0$ in $N_g(x)$ give rise to a
decay of the expectation values like $x_0^{-3}$ or faster, and
the asymptotically leading contributions due to the odd powers
are linear in $\phi_0$.

Let us summarize at this point the results of the preceding
analysis: the algebraic structure of the asymptotic thermal
field reproduces correctly the general asymptotic form of
the thermal correlation functions, irrespectively of the
underlying interaction. The fact that different interactions
lead to different asymptotic thermal correlation functions
manifests itself in the degeneracy of the KMS states on ${\cal A}_0$.
Field equations can be formulated
in an asymptotic sense and lead
to constraints on the form of the KMS states.
As we shall see in the next section, it is then a matter of
some elementary computations to determine their concrete form.

\section{Computation of asymptotic propagators}
\setcounter{equation}{0}
In order to illustrate the preceding general results, let us
consider first a theory where the field $\phi_0$ satisfies an asymptotic
field equation fixed by
\begin{equation} \label{example}
N(x) = \square \, \phi_0 (x) + m_0^2 \, \phi_0 (x) + g \, \phi_0^3 (x)
\end{equation}
with $g < 0$ (corresponding to a quartic interaction with negative
coupling; this interaction is known to be asymptotically free
\cite{Sy}). In view of the remarks after the preceding definition and
the fact that the KMS states are invariant under
space-time translations, it suffices for the determination of
KMS states which are compatible with the asymptotic
field equation given by $N(x)$ to consider the condition
\begin{equation} \label{ascon}
\lim_{x_0 \rightarrow \infty} \,
x_0^{3/2} \, \obf{N(x) \, \phi_0(0)} = 0.
\end{equation}
Here we have omitted the convolution with $g$
since $N(x)$ contains only odd powers of $\phi_0$.
Making use of relations (\ref{freekms}) and
(\ref{cube}), we obtain
\begin{eqnarray}
& & \obf{N(x) \, \phi_0(0)} =
\big( \square + m_0^2 + 3 g k (\beta) \big) \,
 \obf{Z(\bx)} \,  W_{m_0,\beta} (x) \\
& & = \big((- \Delta + 3 g k (\beta)) \, \obf{Z(\bx)} \big) \,
W_{m_0,\beta} (x)
- 2 \, \nabla  \, \obf{Z(\bx)} \cdot \nabla  \, W_{m_0,\beta} (x).
\nonumber
\end{eqnarray}
Now the distribution $\nabla  \, W_{m_0,\beta} (x)$ decreases like
$x_0^{-2}$ for asymptotic $x_0$, provided $\bx$ varies within compact
sets
(\cf Appendix A).
Hence in view of the $x_0^{-3/2}$ decay of $ W_{m_0,\beta} (x)$, 
condition (\ref{ascon}) is satisfied if and only if
\begin{equation} \label{stateq}
(- \Delta + 3 g k (\beta)) \, \obf{Z(\bx)} = 0.
\end{equation}
Restricting attention to KMS states which are invariant under
rotations in their rest systems
and taking into account the normalization of
$ \obf{Z(\bx)}$, the unique solution of the latter equation is, setting
$\kappa(\beta) = (3 \, |g| \, k (\beta))^{1/2}$,
\begin{equation}  \label{Z1}   
\obf{Z(\bx)} = {\sin \big( \kappa(\beta) |\bx | \big) \over \kappa(\beta)
|\bx |} \ .
\end{equation}
So, for any given temperature, the two-point functions (\ref{freekms})
of the KMS states and hence the states itself
are completely fixed by the asymptotic interaction.
It is of interest to have a closer look at the properties of these
functions,
\begin{equation} \label{result1}
\obf{\phi_0(x_1) \phi_0(x_2)} =
{\sin \big( \kappa(\beta) |\bx_1 - \bx_2 | \big) \over
\kappa(\beta) |\bx_1 - \bx_2 |} \
W_{m_0,\beta} (x_1 - x_2).
\end{equation}
We first notice that the functions converge to the
two-point function of a non-interacting field at given
inverse temperature $\beta$ in the limit $g \nearrow 0$,
as one might expect in an asymptotically free theory.
The quantity $\kappa(\beta)^{-1}$ can be interpreted as
mean free path of the constituent particles of the
state. It tends to $\infty$ for small $g$ and small temperatures
and behaves like $2 |g|^{-1/2} \beta$ for very large temperatures.
If the spatial coordinates of the fields
have a distance which is small compared
to $\kappa(\beta)^{-1}$, the two-point function coincides
with that of the non-interacting situation.
For larger distances, the function exhibits
patterns of a standing wave, consisting of a
superposition of an incoming and an outgoing spherical
wave whose amplitudes decrease with time.

The momentum space spectral density, obtained from the
two-point function by Fourier transformation, is given by
\begin{equation} \label{specdens} 
\widetilde{\rho}_- (p)
= {1 \over 1 \! - \! e^{-\beta p_0}} \
{\varepsilon(p_0) \over 8\pi \kappa(\beta) |\bp|} \
\theta(p^2 \! - \!  m_\sharp^2\!  + \! 2 \kappa(\beta) |\bp| ) \
\theta(- p^2 \! + \! m_\sharp^2 \! + \! 2 \kappa(\beta) |\bp| ) 
\end{equation}
where $m_\sharp^2 = m_0^2 + \kappa(\beta)^2$.
It is non-negative, in accordance with the
requirement that the functional (\ref{result1})
ought to define a state on ${\cal A}_0$; note that this
feature is a consequence of the asymptotic dynamics.
The support of the spectral density is confined to the region
\begin{equation} \label{supp} 
\big(m_0^2 + (|\bp| - \kappa(\beta))^2 \big){}^{1/2}
\leq |p_0| \leq \big(m_0^2 + (|\bp| + \kappa(\beta))^2 \big){}^{1/2}.
\end{equation}
\begin{figure}[h]
\vspace*{-15mm}
\epsfxsize85mm
\hspace{28mm} 
\epsfbox{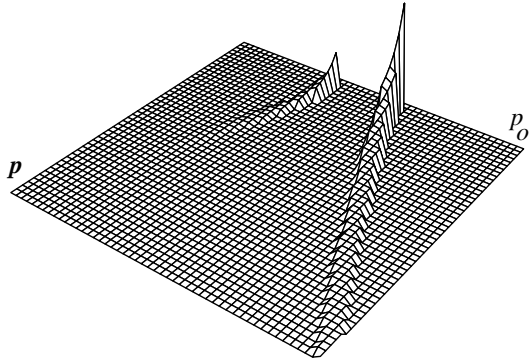}
\caption{Spectral density for negative
coupling.}
\end{figure}
Hence the energy needed to create a constituent particle
of momentum $\bp = 0$ in the thermal background
is equal to $m_\sharp$. It is larger than $m_0$ since one has to inject
in addition to the rest mass of the particle its interaction
energy with the thermal background. An analogous statement
holds for the energy gained by the
removal of a particle from the state (creation
of a hole). For non-zero momenta $\bp$, the mass shell of the
particle, respectively hole, is spread
over a region whose width increases with increasing temperature.
A qualitative picture of the spectral density $\widetilde{\rho}_-$
is given in Fig.\ 1.

The spectral density vanishes for large spacelike momenta,
hence the state complies with the relativistic KMS condition
(which can also be seen directly from Eq.~(\ref{result1})).
It thus has all properties, outlined
in \cite{BrBu3}, which are expected to be characteristic for the dissipative
propagation of the constituents of a thermal state.
By similar elementary
computations, one can also determine the
Fourier transforms of the time-ordered, retarded
and advanced two-point functions from the correlation function
(\ref{result1}); this is done for completeness in Appendix B.  

Let us determine now the asymptotic thermal
correlation functions for a quartic interaction with positive coupling
$g > 0$. There one might be inclined to replace in the
two-point function (\ref{result1}) the sine by its hyperbolic
counterpart. Yet the resulting function is not temperate and
does not give rise to a state on ${\cal A}_0$. This apparent
problem is solved by noticing that the regularity
assumptions, made for the functions $\obf{Z(\bx)}$
in the previous section for simplicity, are no longer valid in this case.
In fact, one may impose Eq.\ (\ref{stateq}) only
for $\bx \neq 0$, and the  physically acceptable solutions are given by
\begin{equation} \label{Z2} 
\obf{Z(\bx)} = \mbox{const} \
{e^{-\mbox{\footnotesize $\kappa(\beta)$} \,
|\mbox{\footnotesize \boldmath$x$}|} \over |\bx |} .
\end{equation}

It is instructive to discuss in a more systematic manner how these
solutions arise. To this end we have to reconsider the definition of the
normal-ordered cube of the field $\phi_0$ in situations where the
functions $\obf{Z(\bx)}$ exhibit a singularity at $\bx = 0$.
Keeping $\varepsilon > 0$ fixed for a moment, we define the
regularized cube
\begin{eqnarray}
& & \phi_0^3 (x \, ; \varepsilon) \\
& &  =  \phi_0(x+\beps_1) \phi_0(x+\beps_2) \phi_0(x+\beps_3) 
   - \! \sum \ovf{\phi_0(x+\beps_j) \, \phi_0(x+\beps_k)} \, \phi_0
(x+\beps_l), \nonumber
\end{eqnarray}
where the spatial vectors $\beps_j$, $j=1,2,3$,
form the corners of an equilateral
triangle of side length $\varepsilon$ with $0$ at its
barycenter and the sum extends over all
partitions of $\{1,2,3\}$ into an ordered pair and a singlet.
Replacing the expression given in (\ref{example}) by its regularized version
\begin{equation} \label{quartic}
N_\varepsilon(x) =
{1 / 3} \, \sum_{j=1}^3
\big(\square \, \phi_0 (x + \beps_j) + m_0^2 \, \phi_0 (x + \beps_j) \big)
+ g \, \phi_0^3 (x \, ; \varepsilon),
\end{equation}
we demand as before that the admissible KMS states
satisfy the corresponding compatibility condition (\ref{constraint})
at non-singular points. In particular,
\begin{equation}
\lim_{x_0 \rightarrow \infty} \,
x_0^{3/2} \, \obf{N_\varepsilon(x) \, \phi_0(0)} = 0
\end{equation}
for $|\bx| > \varepsilon$. Restricting as before our attention to isotropic
KMS states satisfying the normalization condition
$\obf{Z(\beps)} = 1$ for $|\beps| = \varepsilon$, this
condition is fulfilled provided
$\obf{Z(\bx)}$ is, for $\bx \neq 0$, a solution of
\begin{equation}
(- \Delta + 3 g k_\varepsilon (\beta)) \, \obf{Z(\bx)} = 0,
\end{equation}
where
\begin{equation}
k_\varepsilon (\beta) = 2 \,
(2 \pi)^{-3} \int \! dp \ \theta (p_0) \delta (p^2 - m_0^2) \,
(e^{\beta p_0} - 1)^{-1} \, \cos (\bp \beps).
\end{equation}
The temperate, isotropic and normalized solutions are
\begin{equation} \label{approx}
\obf{Z(\bx)} = \varepsilon \,
e^{\, \mbox{\footnotesize $\kappa_\varepsilon (\beta)$}
\mbox{\footnotesize \boldmath$\varepsilon$} } \ \,
{e^{-\mbox{\footnotesize $\kappa_\varepsilon (\beta)$} \,
|\mbox{\footnotesize \boldmath$x$}|} \over |\bx|},
\end{equation}
where $\kappa_\varepsilon (\beta) = (3 \, g \, k_\varepsilon
(\beta))^{1/2}$. They fix the KMS states which are asymptotically
compatible with the regularized interaction.

In order to obtain 
a meaningful result
in the limit $\varepsilon
\rightarrow 0$,  
one has to renormalize
the field $\phi_0$ by the temperature independent
factor $(\kappa_0 \, \varepsilon)^{-1/2}$,
$\kappa_0$ being some fixed energy, so as to compensate the
factor $\varepsilon$ in (\ref{approx}). Since
$\kappa_\varepsilon (\beta) \rightarrow \kappa (\beta)$
for $\varepsilon \rightarrow 0$, the two-point
functions of the renormalized field arising in this limit
are
\begin{equation} \label{result2}
\obf{\phi_0(x_1) \phi_0(x_2)} =
{e^{-\mbox{\footnotesize $\kappa(\beta)$} \,
|\mbox{\footnotesize \boldmath$\bx_1 - \bx_2 $}|} \over
\kappa_0 \, |\bx_1 - \bx_2 |} \
W_{m_0,\beta} (x_1 - x_2).
\end{equation}
These functions exhibit a universal singular
behaviour at coinciding arguments which may be attributed to
the bad ultraviolet properties of the quartic interaction for
positive couplings. In particular, the field no longer
satisfies canonical equal time commutation relations
in the thermal states. Moreover, the two-point functions
do not converge to the thermal correlation functions of a
non-interacting field for $g \searrow 0$. They also do not
approach the vacuum situation in the limit of zero
temperature, but define a ground state. So the physical
situation described by these functions is quite different from
that for negative coupling.

The momentum space spectral density of 
the two-point function (\ref{result2}) is given by the 
following expression 
(see also the corresponding Green's function in 
Appendix B):
\begin{equation} \label{specdens2} 
\widetilde{\rho}_+ (p) =
{1 \over 1 \! - \! e^{-\beta p_0}} \
{\varepsilon (p_0) \over 8 \pi^2 \kappa_0 |\bp|} \ 
\theta(p_0^2 \! - \! m_0^2) \
\ln{\big((p_0^2 - m_0^2)^{1/2} + |\bp|\big)^2 
+ \kappa(\beta)^2 \over
\big((p_0^2 - m_0^2)^{1/2} - |\bp|\big)^2 + \kappa(\beta)^2}.
\end{equation}
It is again non-negative, so also in the case of positive
coupling there exist thermal states on the algebra of asymptotic
thermal fields which are compatible with the asymptotic
dynamics. But these states no longer comply with the relativistic
KMS condition due to their singular behaviour in configuration
space.

In contrast to the previous case, the support of the density
$\widetilde{\rho}_+$ is now spread over the whole region
$|p_0| \geq m_0$. It has, for small temperatures, sharp
maxima in $p_0$ for given $|\bp|$ which define approximate
dispersion laws for the constituent particles and 
the corresponding holes in the thermal background.
Again, one has to inject an energy of order $m_\sharp$
in order to create with substantial probability a particle
of zero momentum in the thermal background.
In Fig.\ 2 a qualitative picture
of the spectral density $\widetilde{\rho}_+$ is given
for the same mass, modulus of the coupling constant, temperature
and range of momenta as in Fig.\ 1.

\begin{figure}[t]
\vspace{-25mm}
\epsfxsize103mm
\hspace{25mm} 
\epsfbox{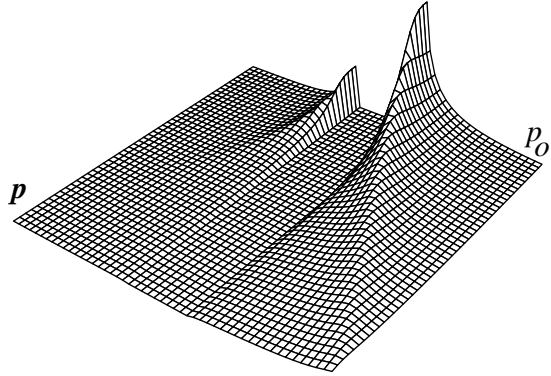}
\caption{Spectral density for positive coupling.}
\end{figure}

The preceding results show that the algebra of
asymptotic thermal fields and the concept of asymptotic field
equations provides an efficient tool for the non-perturbative
computation of the asymptotic thermal propagators. These propagators 
can be computed in a straight-forward manner from the underlying field
equations in which the physical mass $m_0$ of the
particle and strength $g$ of the interaction are encoded.
We mention as an aside that in the perturbative computation
of thermal Green's functions these parameters
can be fixed independently of the temperature \cite{MuKoRe}.

Since there did not appear any inconsistencies in the
present asymptotic analysis, there arises the interesting question
of whether also the full theory with a quartic interaction
can be defined in thermal states, in
contrast to the vacuum, where it is expected
to be trivial for positive couplings and unstable for
negative ones. The preceding results seem to indicate
that free field theory would not be the appropriate starting
point, however, for the construction if $g > 0$.
One should start instead from a theory, where
the specific short distance structure of the propagators
is taken into account from the outset. This is in
accordance with the ideas expounded in \cite{Kl} which
may be of relevance also in the present context.

\section{Conclusions}
\setcounter{equation}{0}
Starting from general physical assumptions,
we have determined the asymptotic form of the thermal correlation
functions of scalar fields. It turned out that discrete mass
contributions in the K\"all\'en-Lehmann type representation
of the two-point functions give rise to asymptotically
leading terms which have a rather simple form: they are products of
the thermal correlation function of a free field and a damping
factor describing the dissipative effects of the model-dependent
thermal background.

The asymptotic correlation functions can be interpreted in terms
of quasifree states on some 
central extension of the algebra of a
free scalar field. Intuitively speaking, this field carries an
additional stochastic degree of freedom which manifests itself
in a central element appearing in the commutation relations and
couples to the thermal backgrounds. In thermal states describing
pure phases, this central element ``freezes'' in the sense that
it attains sharp c-number values, whereas in mixtures of such states it is
statistically fluctuating.

The concrete examples of two-point functions computed in the
present investigation nicely illustrate the conclusions of the
general discussion in \cite{BrBu3}. They corroborate
the idea that the stable ``constituent particles'' forming a thermal
state give rise to discrete mass contributions in the damping
factors. The existence of such discrete contributions has also
been established in the presence of spontaneous symmetry breaking
\cite{BrBu4}, where they can be attributed to the appearance of
Goldstone Bosons. All these results provide evidence to the effect
that the constituents of thermal states produce an unambiguous
signal in the correlation functions in
configuration space. It is not so easy, however, to
discriminate them in momentum space from other types of excitations
since the damping factors wipe out their discrete mass shells.

It is apparent that the present non-perturbative method for the
analysis and computation of the asymptotic form of the thermal
correlation functions can be extended to theories with a more complex
field content. There appears, however, a technical problem in the
treatment of fields affiliated with massless particles. In order to
identify their asymptotically leading contribution in the thermal
correlation functions one needs a more refined analysis which allows
one to distinguish the effects of massless  constituent particles
from those of collective low-energy excitations. We hope to return
to this interesting problem elsewhere.

\begin{appendix}
\section{Asymptotic behaviour of free propagators}
\setcounter{equation}{0}
In this appendix we establish asymptotic bounds on the smeared--out
free thermal correlation functions, used in Sec.~2, which
are uniform in the mass. Let
$h \in {\cal S} (\RR)$ be any test function whose Fourier transform
vanishes at the origin. We consider the functions,
$\bx$ varying in compact sets and $x_0$ being positive,
\begin{eqnarray}
x_0 & \rightarrow &
\int \! dp \, \varepsilon (p_0) \delta(p^2-m^2) \,
(1 - e^{-\fbeta \fpo})^{-1} \,
\widetilde{h} (p_0) \, \bp^{\bfmu} \, e^{\fim (\fpo \fxo -
\bfp \bfx)} \nonumber \\
& = & \sum_{\fsigma = \pm} \sigma \int \! {d\bp \over 2 \omega} \,
(1 - e^{- \fsigma \fbeta \fomega})^{-1} \,
\widetilde{h} (\sigma \omega) \, \bp^{\bfmu} \,
e^{\fim (\fsigma \fxo \fomega - \bfx \bfp)},
\end{eqnarray}
where $\omega = (\bp^2 + m^2)^{1/2}$ and
$\bp^{\bfmu} = p_1^{\mu_1} p_2^{\mu_2} p_3^{\mu_3}$
is any given monomial in the components of $\bp$ of degree
$\mu = \sum_j \mu_j$.
Performing the spherical integration and
making the substitution $w = (\omega - m)/x_0$, we can proceed to
\begin{equation} \label{expr}
\hspace*{0pt} = \ \ \sum_{\fsigma = \pm}
e^{\fim \fsigma \fm \fxo} x_0^{-(\mu + 3)/2} \int_0^\infty \! dw \,
e^{\fim \fsigma \fw} \, w^{(\mu + 1)/2} \, k_\sigma (w/x_0;m),
\end{equation}
where
\begin{equation}
\begin{split}
k_\sigma (v;m) & =  \sigma \, (v + 2m)^{(\mu +1)/2} \,
(1 - e^{- \fsigma \fbeta (\fv + \fm)})^{-1} \,
\widetilde{h} (\sigma (v + m)) \\
& \times  \int \! d\,\Omega(\be) \,
e^{-\fim (\fv^2 + 2 \fm\fv)^{1/2} \bfx \bfe} \,  \be^\bfmu
\end{split}
\end{equation}
and $d\,\Omega(\be)$ denotes the measure on the unit sphere
$|\be|=1$. Since $\widetilde{h} (0) = 0$,
the latter functions are, for fixed $m \geq 0$, continuous and
rapidly decreasing in $v$. Moreover, since $\widetilde{h}$ is a
test function,
we obtain for fixed $0 < \delta < 1/2$ and $n \in \NN$ bounds
on the $n$-fold derivative of $k_\sigma$ of the form
\begin{equation} \label{aa}
\sup_{v > 0} | v^{n-\delta} \, \partial_v^{\, n} \, k_\sigma (v;m) |
\leq c_N \, (1 + m)^{-N},
\end{equation}
which hold uniformly on compact sets in $\bx$ and for any $N \in \NN$.
It is an immediate consequence of this estimate that
\begin{equation} \label{bb}
|k_\sigma (v;m) - k_\sigma (0;m) | \leq \delta^{-1} v^\delta \,
\sup_{u > 0} | u^{1 -\delta} \, \partial_u \, k_\sigma (u;m) |
\leq c_N \, v^\delta (1 + m)^{-N}.
\end{equation}
We now split the integral in (\ref{expr}) into
\begin{equation}
\begin{split}
&  k_\sigma (0;m) \, \lim_{\varepsilon \searrow 0} \,
 \int_0^\infty \! dw \,
 e^{\fim \fsigma \fw - \fepsilon \fw} \, w^{(\mu + 1)/2}  \\
& + \lim_{\varepsilon \searrow 0} \,
\int_0^\infty \! dw \,
e^{\fim \fsigma \fw - \fepsilon \fw} \, w^{(\mu + 1)/2}
\, (k_\sigma (w/x_0;m) - k_\sigma (0;m)).
\end{split}
\end{equation}
The first term
is a continuous, rapidly decreasing function of $m$.
For the modulus of the second term we obtain by
$\nu$-fold partial integration, $(\mu + 3)/2 < \nu \leq (\mu + 5)/2$,
an upper bound of the form
\begin{equation}
\begin{split}
& \Big| \int_0^\infty \! dw \, (e^{\fim \fsigma \fw - \fepsilon \fw} - 1) \,
w^{(\mu + 1)/2 - \nu} \,
\sum_{n=0}^\nu c_n \, (w /x_0 )^{n} \,
(\partial^n k_\sigma) (w/x_0;m) \Big| \\
& \leq c_N \, |x_0|^{-\delta} \, (1+m)^{-N} \,
\int_0^\infty \! dw \,
|e^{\fim \fsigma \fw - \fepsilon \fw} - 1| \, w^{(\mu + 1)/2 - \nu + \delta},
\end{split}
\end{equation}
where $(\partial^0 k_\sigma) (v;m) \doteq
k_\sigma (v;m) - k_\sigma (0;m)$ and we made use of the
bounds (\ref{aa}) and (\ref{bb}). As
$1 < \nu - (\mu + 1)/2 - \delta < 2$, the latter integral
exists and converges in the limit
$\varepsilon \searrow 0$. So, to summarize, we find that there holds
\begin{equation}
h * \partial^\bfmu_\bfx \, \wfm (x)
= x_0^{-(\mu + 3)/2} \big( e^{-imx_0} k_+(m) +
e^{imx_0} k_-(m) \big) + r(m,x),
\end{equation}
where $k_\pm (m)$ are continuous functions which decrease
faster than any inverse power of $m$ in the limit of large $m$.
The remainder $r(m,x)$ is continuous in $m$ and
bounded by
\begin{equation}
|r(m,x)| \leq c_N \, x_0^{-(\mu + 3)/2 - \delta} \,  (1 + m)^{-N},
\end{equation}
uniformly on compact sets in $\bx$. By similar arguments one
obtains an analogous result for large negative $x_0$.
We note in conclusion that the preceding statements hold for
fixed mass $m > 0$ also without the time-smearing by
the test function $h$.

\section{Asymptotic thermal Green's functions}
\setcounter{equation}{0}
In the main text we have focussed our attention on the  
correlation functions of the thermal fields.
As the standard real and imaginary time formalisms of thermal 
quantum field theory are based on the time-ordered, respectively
retarded and advanced functions, we translate here our 
results into these perhaps more familiar
settings. We begin by recalling the relations between these 
functions and then turn to the concrete examples discussed in Sec.\ 5.

Taking into account the anticipated invariance of the thermal
states under space-time translations, the thermal correlation functions 
\begin{equation} \label{W} 
W(x_1-x_2) \doteq
\ob{\phi(x_1) \phi(x_2)}
\end{equation}
fix the commutator function $C$ by the formula
\begin{equation} \label{C} 
C(x) = W(x) - W(-x).
\end{equation}
Because of locality, $C$ has support in the union of the closed 
forward and backward lightcones $\overline{V}^\pm$. The retarded
and advanced functions $R$, $A$ have support in $\overline{V}^+$
and $\overline{V}^-$, respectively, and are obtained by
splitting $i \, C$ into 
\begin{equation} \label{Cra} 
 i \, C(x) = R(x) - A(x),
\end{equation}
the factor $i$ being a matter of convention. In the cases of interest 
here, this splitting can be performed without ambiguities in view
of the mild singularities of $C$ at the origin. The time-ordered
and anti-time-ordered 
functions $T$ and $\overline{T}$, respectively, are obtained
by setting 
\begin{eqnarray} \label{tau}
T(x) & = & W(x) -i A(x) = W(-x) -i R(x) \\ 
\label{taubar} 
{\overline T}(x) & = & W(x) +i R(x) = W(-x) +i A(x).  
\end{eqnarray}

We now pass to Fourier-space. It is an interesting feature of   
the asymptotic thermal fields considered in Sec.\ 4 
that the Fourier transforms 
$\widetilde{C}$ of the thermal 
commutator functions vanish in the region ${\cal{R}}_{m_0} =
\{ p \in \RR^4 : |p_0| < m_0 \}$, \cf the 
examples considered in Sec.\ 5. In this situation one can
establish the existence of Green's functions $\widetilde{G}$
which are analytic 
in momentum space by the following standard argument: 
The Fourier transforms $\widetilde{R}$, 
$\widetilde{A}$ of the retarded and advanced functions can be
extended analytically into the forward and backward tubes
${\cal{T}}^\pm = \{ k \in \CC^{\,4} : 
\mbox{Im} \, k \in V^\pm \}$, respectively,
because of their support properties in configuration space. 
Moreover, from the momentum space version of (\ref{Cra}),
\begin{equation}  
\widetilde{R}(p) - \widetilde{A}(p) = i \, \widetilde{C}(p),
\end{equation}
one sees that $\widetilde{R}$ and $\widetilde{A}$ coincide on  
${\cal{R}}_{m_0}$. It thus  
follows from the Edge-of-the-Wedge theorem that $\widetilde{R}$, 
$\widetilde{A}$ are pieces of an analytic function $\widetilde{G}$,
the Green's function, with domain of holomorphy containing 
$ {\cal{T}}^+ \cup {\cal{T}}^- \cup {\cal{R}}_{m_0}.$
In particular, there holds the discontinuity formula
\begin{equation}  
\lim_{\varepsilon \searrow 0} \widetilde{G}(p_0 + i \varepsilon,\bp) - 
\lim_{\varepsilon \searrow 0} \widetilde{G}(p_0 - i \varepsilon,\bp)
= i \widetilde{C}(p).
\end{equation}
For real spatial momenta $\bp$, 
the Green's function $\widetilde{G}$ can thus be obtained by the 
Cauchy integral
(or dispersion relation) on the twofold cut complex $k_0$-plane,
\begin{equation} \label{disp} 
\widetilde{G}(k_0, \bp)= 
{1\over 2\pi}\ \int dp_0 \   
{{\widetilde C}(p_0, \bp) \over p_0 - k_0 }.  
\end{equation} 
Although this formula does not exhibit by itself the analytic 
continuation of $\widetilde G$ in all momentum space variables,
it is of practical use for computations 
in the specific examples treated below.  
Moreover, since $p_0 \, \widetilde{C}(p)$ is positive
as a consequence of the KMS condition and the positivity
of $\widetilde{W}(p)$,  
it shows that $\widetilde{G}(p)$ is positive on    
the coincidence region ${\cal R}_{m_0}$.    
The latter property distinguishes the restriction
of the analytic Green's function
to its physical-sheet domain based on
$ {\cal{T}}^+ \cup {\cal{T}}^- \cup {\cal{R}}_{m_0}$.

The Fourier transforms of the 
(anti) time-ordered functions are given in terms of 
${\widetilde R}$ and  $ {\widetilde A}$
(and hence in terms of the boundary values of $\widetilde{G}$)   
by the formulas 
\begin{eqnarray} \label{Ftau} 
\widetilde{T} (p) & = & 
i \, {\widetilde{A}(p) - \widetilde{R}(p) \over  1-e^{-\beta p_0}}
-i \, {\widetilde{A}(p)} \\  
\label{Ftaubar} 
\widetilde {\overline T}(p) & = & 
i \, {\widetilde{A}(p) - \widetilde{R}(p) \over  1-e^{- \beta p_0}}
+i \,  {\widetilde{R}(p)}. 
\end{eqnarray}
We note that in the region of coincidence ${\cal R}_{m_0}$
these equations reduce (as in the vacuum case) to 
\begin{equation} \label{coincid} 
{\widetilde G}(p) = {\widetilde R}(p) = {\widetilde A}(p) = i \widetilde T (p) 
= -i \widetilde {\overline T}(p).  
\end{equation}

After these preliminaries, we can now turn to the two concrete 
examples treated in Sec.\ 5. In view of the preceding formulas,
relating the functions of interest to the Green's function
in a straightforward manner, it suffices to determine the latter.

In the case of negative coupling, $g < 0$, the Fourier transform
of the commutator function, \ie the spectral function, 
is the odd part of the spectral density 
(\ref{specdens}), multiplied
by $2$. Thus it is given by 
\begin{equation} \label{tildeC1} 
\widetilde C_- (p_0,\bp)  
= {\varepsilon(p_0) \over 8\pi \kappa(\beta) |\bp|} \,
\theta(p^2 \! - \!  m_\sharp^2\!  + \! 2 \kappa(\beta) |\bp| ) \
\theta(- p^2 \! + \! m_\sharp^2 \! + \! 2 \kappa(\beta) |\bp| ), 
\end{equation}
where $m_\sharp^2 = m_0^2 + \kappa(\beta)^2$.
By performing the above Cauchy integral one obtains the 
corresponding Green's function for complex $(k_0,\bk)$, 
\begin{equation} \label{disp1}
\widetilde{G}_- (k_0,\bk)   =
 {1\over 16\pi^2 \kappa(\beta) (\bk^2)^{1 / 2} } \,
\ln {k^2 \! - \!  m_\sharp^2\!  - \! 2 \kappa(\beta) (\bk^2)^{1 / 2} 
\over k^2 \! - \!  m_\sharp^2\!  + \! 2 \kappa(\beta) (\bk^2)^{1 / 2} },
\end{equation}
where the principal (positive) branch of the logarithm 
and $(\bp^2)^{1 / 2} = |\bp|$ have to be chosen in the region 
${\cal R}_{m_0}$ of the physical sheet in view of 
the above positivity constraint on $\widetilde{G}_-$.
The logarithmic singularity of this holomorphic function is carried by the 
complex hypersurface $\Sigma_{\sharp}$ with equation 
$ (k^2 - m_\sharp^2)^2 - 4\kappa(\beta)^2 \bk^2 =0 $ which 
can explicitly be seen not  
to intersect the domains $\cal{T}^+$ and $\cal{T}^-$,
while its trace on the reals is the border of the region 
described by (\ref{supp}). The set with equation   
$\bk^2 =0$ does not carry singularity in the physical sheet and at $\bk =0$
one obtains  
\begin{equation} 
\widetilde{G}_- (k_0,0)   =
 - {1\over 4\pi^2 } \,
{1\over k_0^2\  -\  m_\sharp^2}. 
\end{equation}
The retarded, advanced and (anti) time-ordered 
functions are obtained from $\widetilde{G}_-$ by the preceding formulas.
We note that the coincidence relation (\ref{coincid}) holds not only on 
${\cal R}_{m_0}$, but everywhere in the complement of the region 
(\ref{supp}), yielding an extended physical sheet domain for 
$\widetilde{G}_-$. 

Turning to the case of positive coupling, $g >0$, we see from 
relation (\ref{specdens2}) that the corresponding 
commutator (spectral) function is equal to 
\begin{equation}
\widetilde{C}_+ (p_0,\bp) =
{\varepsilon(p_0) \over 8\pi^2 \kappa_0 \, |\bp|} \,
\theta(p_0^2 \! - \! m_0^2) \,
\ln {\big( (p_0^2 - m_0^2)^{1/2} + |\bp| \big)^2 + \kappa(\beta)^2 \over
\big( (p_0^2 - m_0^2)^{1/2} - |\bp| \big)^2 + \kappa(\beta)^2}.
\end{equation}
Inserting this function into (\ref{disp}), 
splitting the logarithm  
into two parts and making a contour-distortion argument 
now yields the analytic Green's function 
\begin{equation} \label{Gat}
\widetilde{G}_+ (k_0,\bk)   =
 {i\over 8\pi^2 \kappa_0 \, (\bk^2)^{1 / 2} } \,
\ln{(k_0^2 - m_0^2)^{1 / 2} + (\bk^2)^{1 / 2} + i\kappa(\beta) 
\over (k_0^2 - m_0^2)^{1 / 2} - (\bk^2)^{1 / 2} + i\kappa(\beta)}, 
\end{equation}
where in the region ${\cal R}_{m_0}$  
one has to choose again the principal branch of the logarithm,
$(\bp^2)^{1 / 2} = |\bp|$ and
$(p_0^2 - m_0^2)^{1 / 2} = i \, |p_0^2 - m_0^2|^{1 / 2}$.

The logarithmic singularity of $\widetilde{G}_-$ is now 
carried by the complex hypersurface  $\Sigma_{\flat}$ with equation 
$ (k^2- m_\flat^2)^2 + 4\kappa(\beta)^2 \bk^2 =0 $,
where $ m_\flat^2 = m_0^2 - \kappa(\beta)^2$. 
In contrast to $\Sigma_{\sharp}$, the surface 
$\Sigma_{\flat}$ 
intersects the domains $\cal{T}^\pm$,
but the corresponding points appear as singularities of 
$\widetilde{G}_+$ only in unphysical sheets. 
If $m_{\flat}^2$ is negative, \ie at large temperatures,
$\Sigma_{\flat}$ does not contain any real points. For 
low temperatures   
the only real points of $\Sigma_{\flat}$ are 
$\pm (m_{\flat}, \bO)$; they appear as second sheet singularities
(``anti-boundstates''), cf.\ Eq. (\ref{Gat0}) below.   

On the real border of its physical sheet domain, the only singularity 
of $\widetilde{G}_+$ 
is the square-root type threshold singularity  
at $p_0 = \pm m_0$ for all $\bp$ since,  
here again, the set with equation   
$\bk^2 =0$ does not carry singularity in the physical sheet.  
At $\bk =0$, one obtains
\begin{equation} \label{Gat0} 
\widetilde{G}_+ (k_0,0)   =
{i\over 4\pi^2 \kappa_0}\   
{1\over (k_0^2 - m_0^2)^{1 / 2} + i\kappa(\beta)}. 
\end{equation}

The retarded and advanced functions,  
defined in terms of the corresponding boundary 
values of $\widetilde{G}_+$, and the (anti) time-ordered
functions given by relation (\ref{Ftau})
now satisfy the coincidence
relation (\ref{coincid}) only in the region ${\cal R}_{m_0}$.

\end{appendix}

\vskip 0,4cm 
\noindent {\bf \Large Acknowledgements} \\[2mm]
The authors are grateful for hospitality and 
financial support by the Universit\"at G\"ottingen 
and the CEA - Saclay, respectively.


\begin{thebibliography}{99}
\bibitem{NaReTh}
H.~Narnhofer, M.~Requardt and W.~Thirring,
Commun.\ Math.\ Phys.\  {\bf 92}, 247 (1983).

\bibitem{La}
N.P.~Landsman,
Annals Phys.\  {\bf 186}, 141 (1988).

\bibitem{St}
O.~Steinmann,
Commun.\ Math.\ Phys.\  {\bf 170}, 405 (1995)

\bibitem{We}
H.A.~Weldon,
Quasiparticles in finite-temperature field theory, preprint hep-ph/9809330

\bibitem{Abri}
A. A. Abrikosov,  
Phys.\ Atom.\ Nucl.\ {\bf 59}, 352  (1996) 
(from Yad.\ Fiz.\ {\bf 59N2}, 372 (1996))  

\bibitem{BrPi}
E.~Braaten and R.D.~Pisarski,
Nucl.\ Phys.\ B {\bf 337}, 569 (1990).

\bibitem{BlI}
J.P. Blaizot and E. Iancu, Phys. Rev. D {\bf 55}, 973 (1997); 
{\bf 56}, 7877  (1997).

\bibitem{Lebe}
M. Le Bellac,  {\sl Thermal Field Theory}. 
Cambridge University Press, Cambridge, England, 1996. 

\bibitem{ArYa}
P.~Arnold and L.G.~Yaffe,
Phys.\ Rev.\ D {\bf 57}, 1178 (1998)

\bibitem{BrBu2}
J.~Bros and D.~Buchholz,
Annales Poincar\'e Phys.\ Theor.\  {\bf 64}, 495 (1996)

\bibitem{LaWe}
N.~P.~Landsman and C.~G.~van Weert,
Phys.\ Rept.\  {\bf 145}, 141 (1987) 

\bibitem{DoJa}
L.~Dolan and R.~Jackiw,
Phys. Rev. D {\bf 9}, 3320 (1974).

\bibitem{BrBu1}
J.~Bros and D.~Buchholz,
Nucl.\ Phys.\ B {\bf 429}, 291 (1994)

\bibitem{BrBu3}
J.~Bros and D.~Buchholz,
Z.\ Phys.\ C {\bf 55}, 509 (1992).

\bibitem{We2}
H.A.~Weldon,
Asymptotic space-time behaviour of HTL gauge propagator,
preprint hep-ph/0009240.

\bibitem{Sy}
K.~Symanzik,
Commun.\ Math.\ Phys.\  {\bf 18}, 227 (1970) and
Commun.\ Math.\ Phys.\  {\bf 23}, 49 (1971).

\bibitem{MuKoRe}
C.~Kopper, V.F.~M\"uller and T.~Reisz,
Annales Henri Poincar\'e {\bf 2}, 387 (2001)

\bibitem{Kl}
J.R.~Klauder,
Self-interacting scalar fields and (non-)triviality,
p. 87 in: {\it Mathematical Physics Towards the 21st Century.
Proceedings Beer-Sheva 1993}, R.N.~Sen and A.~Gersten Eds.
Ben-Gurion University of the Negev Press, 1994

\bibitem{BrBu4}
J.~Bros and D.~Buchholz,
Phys.\ Rev.\ D {\bf 58}, 125012 (1998)

\end{thebibliography}
\end{document}